\documentclass{JHEP3}

\usepackage{epsfig}

\def\hepth#1{ {\tt hep-th/#1}}

\def\be{\begin{equation}}
\def\ee{\end{equation}}
\def\bes{\begin{equation*}}
\def\ees{\end{equation*}}

\def\beqa{\begin{eqnarray}}
\def\beqas{\begin{eqnarray*}}
\def\eeqa{\end{eqnarray}}
\def\eeqas{\end{eqnarray*}}
\def\bea{\begin{eqnarray}}
\def\eea{\end{eqnarray}}
\def\pa{\partial}
\def\nn{\nonumber}

\def\gsim{\;\raise0.3ex\hbox{$>$\kern-0.75em\raise-1.1ex\hbox{$\sim$}}\;}
\def\lsim{\;\raise0.3ex\hbox{$<$\kern-0.75em\raise-1.1ex\hbox{$\sim$}}\;}

\def\La{\Lambda}

\def\F{{\cal F}}

\def\bemat{\left(\begin{array}}
\def\enmat{\end{array}\right)}

\title{{Prepotential and Instanton Corrections in ${\cal N}=2$ 
Supersymmetric $SU(N_1)\times SU(N_2)$ Yang
Mills Theories}}

\author{Marta~G\'omez-Reino \\
DAMTP, Centre for Mathematical Sciences\\
Wilberforce Road, Cambridge CB3 0WA UK \\
E-mail: \email{m.gomez-reino@damtp.cam.ac.uk} }

\abstract{
In this paper we analyse the 
non-hyperelliptic Seiberg-Witten curves derived from M-theory 
that encode the low energy solution of ${\cal N}=2$ 
supersymmetric theories with product gauge groups. 
We consider the case of a $SU(N_1)\times SU(N_2)$ gauge theory
with a hypermultiplet in the bifundamental representation 
together with matter in the fundamental representations 
of $SU(N_1)$ and $SU(N_2)$. By means of the Riemann
bilinear relations that hold on the Riemann surface defined by the
Seiberg--Witten curve, we compute the logarithmic derivative of 
the prepotential with respect to the quantum scales of both gauge groups. 
As an application we develop a method to compute recursively the
instanton corrections to the prepotential in a straightforward way. 
We present explicit formulas for up to third order on both quantum scales. 
Furthermore, we extend those results to $SU(N)$ gauge theories with 
a matter hypermultiplet in the symmetric and antisymmetric 
representation. We also present some non-trivial checks of our
results. 
\vskip0.4cm \hfill\break
\leftline{DAMTP-2003-1, hep-th/0301105}

\keywords{${\cal N}=2$ Supersymmetric Gauge Theories,
Non-hyperelliptic Riemann surfaces, Instanton Corrections} }

\preprint

\begin{document} 

\section{Introduction}

It is by now a well-known fact that, as long as ${\cal N}=2$ supersymmetry is
unbroken, the low energy effective action is given in terms of a holomorphic
prepotential $\F$. The Seiberg--Witten solution for this low-energy 
effective action \cite{sei1} allows us, in principle, to reconstruct 
the prepotential
of the theory using a set of algebro-geometric data. Such solution 
is given in terms of a suitable Riemann surface or algebraic curve $\Sigma$,
 and a preferred meromorphic 1-form $dS_{SW}$, which is known as
Seiberg--Witten differential. This differential induces a special geometry on
 $\Sigma$, and its periods give the spectrum of BPS states of the
theory. Interestingly enough, this solution 
displays remarkable nonperturbative phenomena such as quark confinement by
monopole condensation, when a mass term that breaks supersymmetry
down to ${\cal N} = 1$ is included. 

The original work of Seiberg and Witten \cite{sei1} was developed
for ${\cal N}=2$ theories with gauge group $SU(2)$ with and without
matter in the fundamental representation. Nevertheless, this solution 
was soon extended 
to other gauge groups and matter content by
determining both the appropriate complex curve and meromorphic
differential \cite{sun,klemm,sunmat,hananoz,son}, thus leading to a
substantial progress in our understanding of ${\cal N}=2$ supersymmetric
gauge theories. In fact, the appearance of an auxiliary Riemann
surface made it possible to identify remarkable connections. In
particular, it pointed out the connection
with string theory, where such Riemann surfaces have a concrete physical
meaning. Geometrical engineering \cite{geomeng} as well as 
M--theory/type IIA methods \cite{witten,otro,lopez2}, where a 
fivebrane is wrapped over a Riemann surface such that the theory living in
the flat four dimensional part of the fivebrane become a four
dimensional gauge theory, have greatly enlarged the 
Seiberg--Witten curves that can be found. 
 Therefore, those connections have also enlarged the 
solutions of ${\cal N}=2$ gauge theories that can be studied.

Once the appropriate Riemann surface or algebraic curve is found 
for a given theory the order parameter of the theory, $a_i$,
and its duals, $a_D^i$, are defined as the period integrals of the 
Seiberg--Witten differential defined over the Riemann surface. 
This is done in such a way that the prepotential of
the theory is implicitly defined as $a_D^i=\frac{\pa {\cal F}}{\pa
a_i}$. Then, once one finds the appropriate Riemann 
surface or algebraic curve for a given theory,
the goal is to compute the period integrals and
 integrate them to find the prepotential ${\cal F}(a)$. 

For classical groups, with gauge multiplet and  $N_f$ 
hypermultiplets in the fundamental representation 
the Seiberg--Witten curves encoding the solution of the theory 
are all hyperelliptic \cite{sun,klemm,sunmat,hananoz,son}, ({\it
i.e.} those curves have the form $y^2+P_1(x)y+P_2(x)=0$). 
Nevertheless, for example for $SU(N_1)\times SU(N_2)$
or ${SU}(N)$ with matter in the symmetric or antisymmetric
representation, 
the appropriate curves are non--hyperelliptic, but are cubic 
\cite{witten,otro,lopez2} ({\it i.e.} of the form 
$y^3+P_1(x)y^2+P_2(x)y+P_3(x)=0$). 
Therefore, the problem that appears to compute the 
prepotential of the theory is how to evaluate the period integrals 
for non-hyperelliptic curves. 
A considerable effort has been done in this direction using a 
perturbation expansion of the non-hyperelliptic curve around
its hyperelliptic approximation \cite{sch1}--\cite{sch4}. 
Nevertheless, the computation of 
the dual periods using that method gets very complicated 
and just allows one to compute the first instanton correction to the
prepotential.

In that sense, it is
always useful to find a method that let us determine the form of
${\cal F}$ without going through the actual computation of the
periods. For ${\cal N}=2$ theories with classical gauge
groups and matter in the fundamental representation 
such methods were developed in \cite{hoker2,chan}, expressing the 
logarithmic derivative of the prepotential with respect to the 
quantum scale of the gauge theory in terms
of the moduli of the curve \cite{hoker2,matone,eguchi,sonn}. Also similar 
methods to compute
recursively the instanton corrections to the prepotential were 
developed in \cite{mas1,mas2} using the connections of ${\cal N}=2$ 
theories with integrable systems. 
Inspired by this fact, in this paper we develop a method 
to compute the instanton correction to the prepotential recursively
without computing the dual periods. In particular, we find the 
logarithmic derivatives of the
prepotential for the non-hyperelliptic curves under study and 
use them to compute the prepotential avoiding the actual computation of the
dual periods of the Seiberg--Witten differential. This method is
the non-hyperelliptic generalization of the work done in 
\cite{hoker2,chan,hoker1} for hyperelliptic curves, and 
substantially simplifies the calculations 
done in \cite{sch1}--\cite{sch4}. Furthermore, this simplification
allows us to compute recursively the instanton corrections to the 
prepotential in a remarkably straightforward way.

The structure of the paper is as follows: In the next section we
review the form of the Seiberg--Witten curves for ${\cal
N}=2$ supersymmetric gauge theory with gauge group $SU(N_1)\times
SU(N_2)$, that are derived from M-theory considerations. We
also study the form of those curves and analyse the information that
we can extract from them. In section 3 we calculate the
logarithmic derivatives of the prepotential with respect to the
quantum scales of both groups in terms of the moduli of the curve, 
using the Riemann bilinear relations. In section 4 we develop a 
method to calculate the
instanton corrections to the prepotential recursively using the
previously calculated equation. We also extend, in section 5, the
results of the previous section to the non-hyperelliptic 
curves obtained for $SU(N)$ theories with matter in the symmetric and
antisymmetric representation. Finally, in section 6 we present
the conclusions. 

\section{${\cal N}=2$ supersymmetric $SU(N_1)\times SU(N_2)$
Yang--Mills theories}

In this paper we will focus in a ${\cal N}=2$ supersymmetric 
 $SU(N_1)\times SU(N_2)$ theory 
with one massless hypermultiplet 
in the $(N_1,\bar{N_2})$ bifundamental representation, 
together with $N_{f_1}$ and $N_{f_2}$ matter hypermultiplets 
in the fundamental representation of $SU(N_1)$ and $SU(N_2)$ respectively. 
This theory has a chiral multiplet in the adjoint representation of $SU(N_1)$ 
that contains a complex scalar field $\phi$ 
and a chiral multiplet in the adjoint representation of 
$SU(N_2)$ that contains a complex scalar field $\hat\phi$. 
This theory has a classical potential with flat directions that
 parametrizes the classical moduli space of the theory. Along such flat
 directions $[\phi,\bar{\phi}]$ and $[\hat\phi,\hat{\bar\phi}]$ vanish, and 
the symmetry is broken to $U(1)^{N_1-1}\times U(1)^{N_2-1}$. The low
 energy solution of the theory is encoded in a particular Riemann
 surface that allows us to compute the prepotential of the theory,
 this surface being derived from M-theory considerations.

\EPSFIGURE[t]{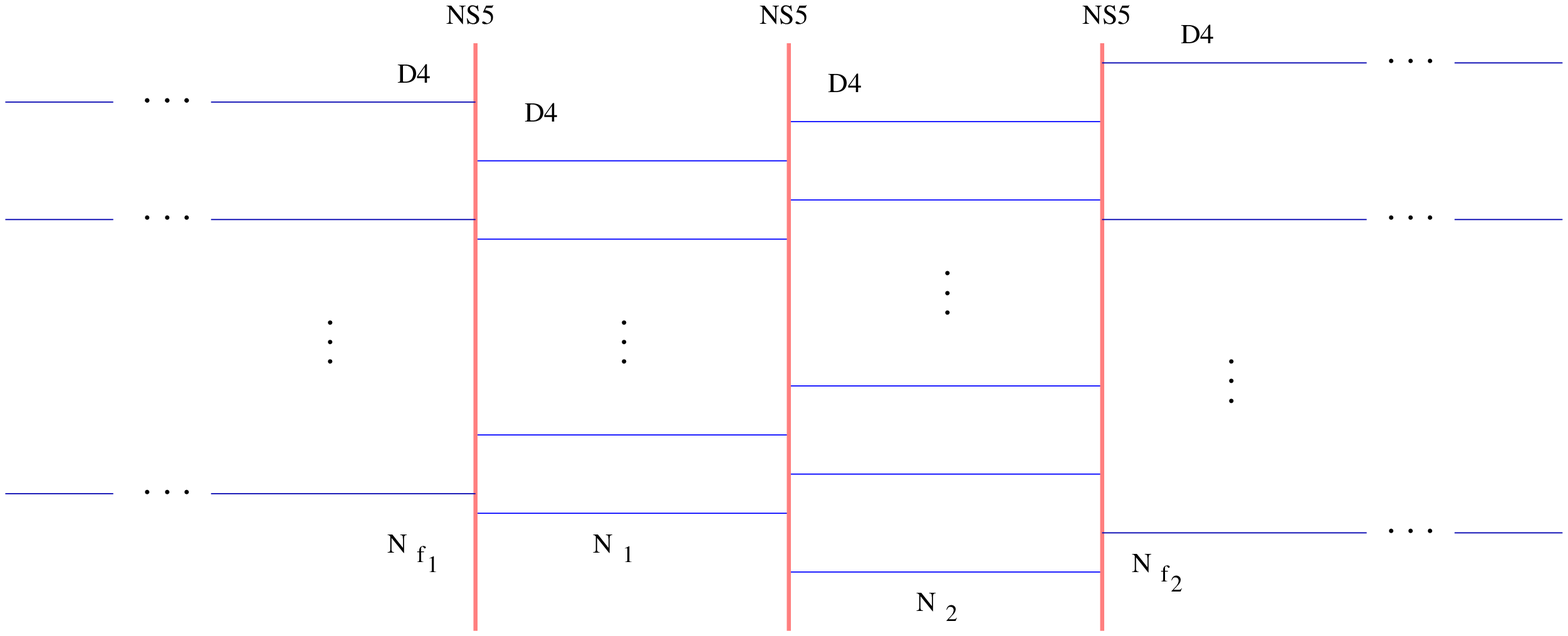, width=.9\textwidth}{Brane picture of
$SU(N_1)\times SU(N_2)$ gauge theories.}

\subsection{Seiberg--Witten curves}

The curve for this theory was derived by Witten in \cite{witten} 
by considering, in type IIA string theory, D4--branes stretched 
between NS fivebranes (see Fig.1). In this context, the Seiberg--Witten  
curve appears when this configuration is lifted to M--theory, as the
configuration becomes a single fivebrane wrapped over a Riemann
surface. This Riemann surface is, in fact, the Seiberg--Witten 
curve of the ${\cal N}=2$ theory. For the theory under study the
Seiberg--Witten curve is given by \cite{witten}
\be
P_0(x)\,y^3\,-\,P_1(x)\,y^2\,+\,\La_1^{\beta_1}P_2(x)\,y\,-\,
\La_1^{2\beta_1}\La_2^{\beta_2}\,P_3(x)\,=0, \label{curve}
\ee
where the coefficients $\beta_1$ and $\beta_2$ are given by
\be
\beta_1\,=\,{2N_1-N_{f_2}-N_1}\,,\,\,\,\,\,\,\,\,\,
\beta_2\,=\,{2N_2-N_{f_1}-N_2}, \label{coef}
\ee
and $\La_1$, $\La_2$ denote the quantum scales of the two gauge groups. 
The requirement of asymptotic freedom, and restriction to the Coulomb 
phase, implies that $\La_1$ and $\La_2$ appear with positive powers 
in (\ref{curve}), that is, $\beta_1$, $\beta_2>0$.

Also in (\ref{curve}) the polynomials $P_1(x)$ and 
$P_2(x)$ denote the characteristic polynomial of $SU(N_1)$ and $SU(N_2)$ 
respectively, and are given by 
\beqa
P_1(x)\,&=&\prod_{i=1}^{N_1}\,(x-e_i)=x^{N_1}-\sum_{k=2}^{N_1}u_k
x^{N_1-k}\label{as}\\
P_2(x)\,&=&\prod_{i=1}^{N_2}\,(x-\hat{e}_i)=x^{N_2}-\sum_{k=2}^{N_2}
\hat u_k x^{N_2-k}\label{b}\,,
\eeqa
and the polynomials $P_0(x)$, and $P_3(x)$ depend just 
on the mass of the hypermultiplets $m_f$, $\hat m_f$, and are given by
\beqa
P_0(x)\,&=&\prod_{i=1}^{N_{f_1}}\,(x+m_i)=x^{N_{f_1}}+
\sum_{k=1}^{N_{f_1}}t_kx^{N_{f_1}-k}\\
P_3(x)\,&=&\prod_{i=1}^{N_{f_2}}\,(x+\hat{m}_i)=x^{N_{f_2}}+
\sum_{k=1}^{N_{f_2}}\hat t_k x^{N_{f_2}-k}\,.
\eeqa

The $(N_1-1)+(N_2-1)$ dimensional moduli space is parametrized classically by 
$e_i$ ($1\leq i\leq N_1$) and $\hat{e}_i$ ($1\leq i\leq N_2$), 
which are the eigenvalues of $\phi$ and $\hat\phi$ respectively, 
and satisfy the constraints $\sum_{i=1}^{N_1} e_i = 0$ and   
$\sum_{i=1}^{N_2} \hat{e}_i = 0$. 
Nevertheless, one should keep in mind that neither 
of these parameters are invariant under Weyl
transformations. On the contrary the symmetric polynomials $u_k$, $\hat
u_k$ in (\ref{as}), (\ref{b}), provide faithful coordinates for 
the moduli space of vacua, so
all the physical quantities extracted from the curve must be given in
terms of those polynomials.

One interesting feature that this Seiberg--Witten curve 
(\ref{curve}) presents is that the map 
\be 
y\rightarrow {\La_1^{\beta_1}\,\La_2^{\beta_2}\over y}
 \label{cambio}
\ee
interchanges the gauge groups, that is, $SU(N_1)
\leftrightarrow SU(N_2)$. This characteristic happens to be very 
useful for the calculations we will describe in the following sections.

The curve (\ref{curve}) is a non-hyperelliptic cubic curve with 
$2N_1+2N_2$ branch points that we will denote
$e_i^\pm$, $i=1,\cdots,N_1$ and $\hat e_j^\pm$, $j=1,\cdots,N_2$. For small
$\La_1$ and $\La_2$, those branch cuts are perturbations of $e_i$ and
$\hat e_i$ respectively. We view then the Riemann surface as 
a three-fold branched covering of the Riemann sphere, with branches one
and two connected by $N_1$ square--root branch cuts joining $e_i^+$
and $e_i^-$ and branches two and three connected by $N_2$ 
square--root branch cuts joining $\hat e_i^+$ and $\hat e_i^-$.

In general terms, a Riemann surface
defined by the relation $F(x,y)=\sum_{i,j}a_{ij}y^ix^j=0$ has a set of 
holomorphic differentials given by $d\omega_{i,j}=
\frac{y^{i-1}x^j}{\pa_x F}dy$, each one of them associated with each one of
the moduli $a_{ij}$ of the curve. In our case, we will have $g
=N_1+N_2-2$ holomorphic differentials associated with each one of the 
moduli $u_k$, $\hat u_k$ \cite{rusos}. For the parametrization 
 of the curve (\ref{curve}) 
we find the following set of holomorphic differentials
\be
d\omega_k=-\frac{yx^{N_1-k}}{\pa_xF}dy
\,,\hspace{2cm} 
d\hat\omega_k=\frac{\La_1^{\beta_1}x^{N_2-k}}{\pa_xF}dy\label{omega}\,.
\ee
Note that (\ref{cambio}), that
interchanges the role of the gauge groups in the Seiberg--Witten curve
also interchanges the set holomorphic differentials $d\omega_k \leftrightarrow 
d\hat\omega_k$, as it should.

The SW differential for these curves is
\be
dS_{SW}=x\frac{{\rm d}y}{y},\label{ds}
\ee
which takes a different value on each one of the three Riemann
branches. This differential is defined in such a way that its
derivative with respect to the moduli of the curve shall give
the holomorphic differentials (\ref{omega}). Note that this condition
is fulfilled in our case as 
$d\omega_k=\partial_{u_k}dS_{SW}$ and $d\hat\omega_k=\partial_
{\hat u_k}dS_{SW}$.

\subsection{Analysis of the curves}

Working with non-hyperelliptic curves, even with cubic ones, is not an
easy task.  The main reason for this is that one needs to 
know the algebraic solution of the curve (that is, $y=y(x)$) in order 
to be able to compute the periods of the
Seiberg--Witten differential. Solutions are generically
not known, unless for cubic curves. But, even in the cubic case, 
they are too complicated to be useful. 
Nevertheless, those solutions can be simplified in the following way: 
If we take the limit $\La_2\rightarrow 0$ on (\ref{curve}) 
we get
\be
P_0(x)y^2-P_1(x)y+\La_1^{\beta_1}P_2(x)=0\label{la2}\,,
\ee
that is, we recover the hyperelliptic curve that
describes the case of a supersymmetric Yang--Mills theory with gauge
group $SU(N_1)$ with $N_{f_1}+N_2$ matter
hypermultiplets in the fundamental representation of the gauge group. 

This can also be seen from the brane picture, as shown in Fig.2. 
Taking the quantum scale $\La_2$ to zero is the same as taking 
the distance between the second and the third
$NS5$--branes in Fig.1 to infinity. 
The gauge coupling of the
$SU(N_2)$ theory then will go to zero and the moduli of the 
theory are seen just as constants with respect to the quantum scale $\La_1$, 
that is, they are seen as matter in the fundamental representation of
$SU(N_1)$. Since the curve (\ref{curve}) obeys the involution (\ref{cambio}), 
the same is true when $\La_1\rightarrow 0$ but 
interchanging the role of the gauge groups. 
Therefore, when $\La_1\rightarrow 0$, we would 
have a $SU(N_2)$ gauge theory with $N_{f_2}+N_1$ matter 
hypermultiplets in the fundamental representation of $SU(N_2)$.

\EPSFIGURE[t]{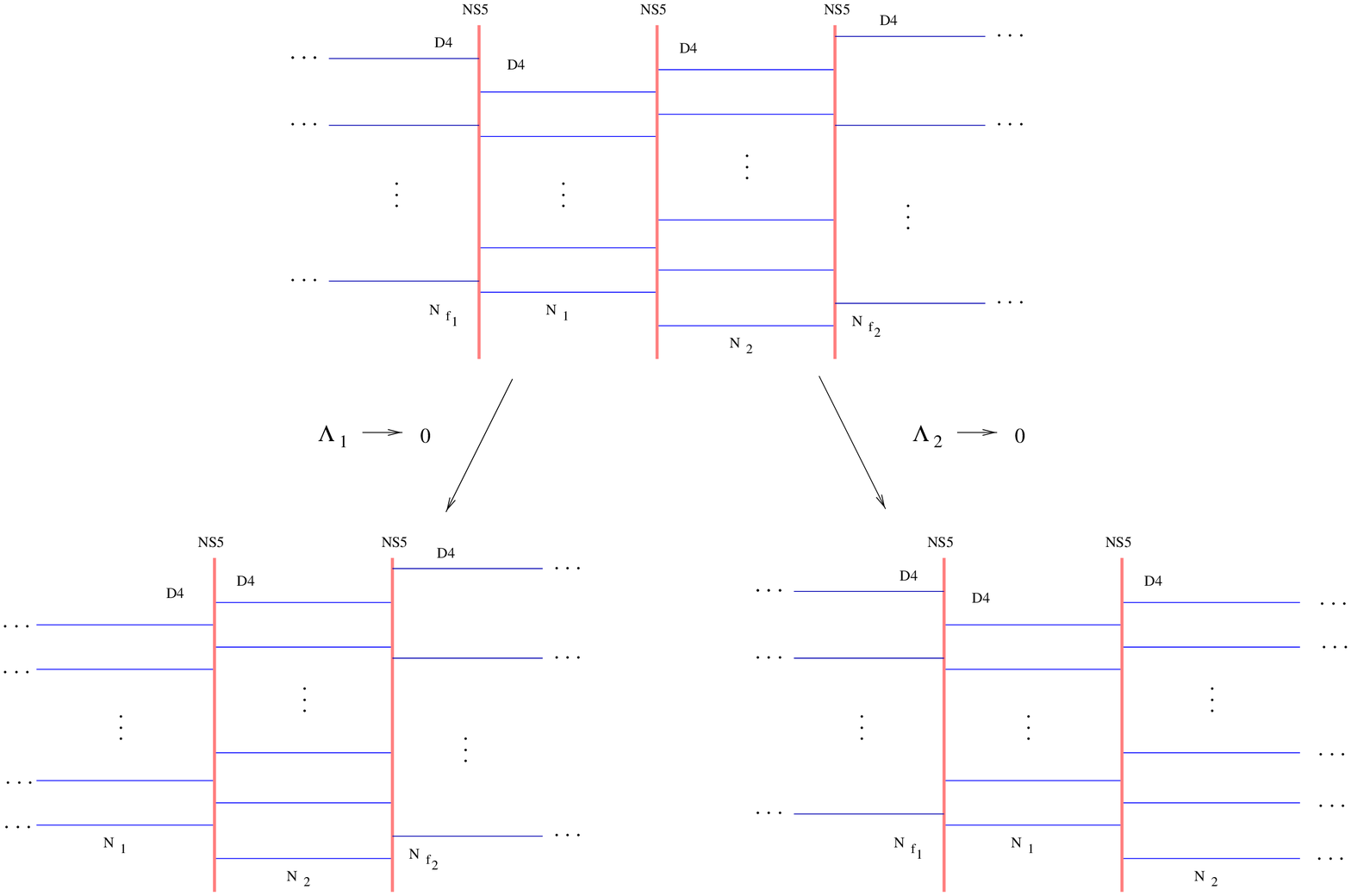, width=.9\textwidth}{Different limits of the theory.}

Then, it is clear that one can obtain the solutions to the
non--hyperelliptic curve (\ref{curve}) 
by means of a systematic series expansion of the curve 
around $\La_2^{\beta_2}=0$, with 
the zeroth-order term being a hyperelliptic curve. This idea was
exploited in \cite{sch1}--\cite{sch4}.  
Using this, the solutions to (\ref{curve}) accurate to the order ${\cal
O} (\La_2^{2\beta_2})$, are
\beqa
y_1&=& \sum_{k=0}^\infty y^{(1)}_k\La_2^{k\beta_2} =\frac{P_1+r}{2P_0}+
{\La_1^{{\beta_1}}P_3(P_1-r)\over {2P_2r}}\La_2^{\beta_2}+\label{app}\\ 
&+&\frac{\La_1^{\beta_1}P_3^2(P_1^4+6\La_1^{2\beta_1}P_0^2P_2^2-
6\La_1^{\beta_1}P_0P_2P_1^2-P_1r^3)}
{2P_2^3r^3}\La_2^{2\beta_2}+\cdots\,,\nonumber\\
y_2&=& \sum_{k=0}^\infty y^{(2)}_k\La_2^{k\beta_2}=
\frac{P_1-r}{2P_0}-{\La_1^{\beta_1}P_3(P_1+r)
\over {2P_2r}}\La_2^{\beta_2}-\\ &-&\frac{\La_1^{\beta_1}P_3^2(P_1^4+6
\La_1^{2\beta_1}P_0^2P_2^2-
6\La_1^{\beta_1}P_0P_2P_1^2+P_1r^3)}
{2P_2^3r^3}\La_2^{2\beta_2}+\cdots\,,\nonumber\\
y_3&=&\sum_{k=1}^\infty
y^{(3)}_k\La_2^{k\beta_2}={\La_1^{\beta_1}P_3\over P_2}
\La_2^{\beta_2}+
{\La_1^{\beta_1}P_1P_3^2\over P_2^3}\La_2^{2\beta_2}+\cdots\,.\label{aprox}
\eeqa
where $r\equiv\sqrt{P_1^2-4\La_1^{\beta_1}P_0P_2}$, and $y_0^{(1)}$ 
and $y_0^{(2)}$ are the solutions to the curve (\ref{la2}), that is,
the hyperelliptic solutions (in the limit $\La_2=0$).
Notice that the involution map (\ref{cambio}) interchanges the
branches $y_1\leftrightarrow y_3$.
 
The perturbative expansion in $\La_2^{\beta_2}$, (\ref{app})--(\ref{aprox}), 
induces a comparable expansion for the SW differential (\ref{ds}). 
For example, in the branch one, using (\ref{app}), we have 
\be
dS_{SW}^{(1)}\,=\left(dS_{SW}^{(1)}\right)_{I}\,+\left(dS_{SW}^{(1)}
\right)_{II}\La_2^{\beta_2}+\left(dS_{SW}^{(1)}
\right)_{III}\La_2^{2\beta_2}+\cdots \label{lambda}
\ee
where 
\be
\left(dS_{SW}^{(1)}\right)_{I}\,=x\frac{d y^{(1)}_0}{y^{(1)}_0}
=\,x\frac{\pa_x\left(\frac{P_1}{(P_0P_2)^{1/2}}\right)}
{\sqrt{\frac{P_1^2}{P_2P_0}-4\La_1^{\beta_1}}}dx +\frac12 x\left(
\frac{\pa_xP_2}{P_2}-\frac{\pa_xP_0}{P_0} \right)dx \,,\label{lambdai}
\ee
Note that this expression is the usual one for the SW 
differential for a hyperelliptic curve, up to terms (the last
term in (\ref{lambdai})) that do not contribute to the periods 
defined in the branch one. The subsequent corrections are 
\beqa
\left(dS_{SW}^{(1)}\right)_{II}&=&-\frac{y^{(1)}_1}{y^{(1)}_0}dx =
-\frac{\La_1^{\beta_1}P_3P_0(P_1-r)}
{P_2r(P_1+r)}\,dx\,.\label{lambdaii}\\
\left(dS_{SW}^{(1)}\right)_{III}&=&-\left(\frac{y^{(1)}_2}{y^{(1)}_0}+
\frac{(y^{(1)}_1)^2}{2y^{(1)}}\right)dx =\frac{2\La_1^{2\beta_1}P_3^2P_0^2
((9P_1-5r)\La_1^{\beta_1}P_0P_2-2P_1^2(P_1-r))}
{P_2^2r^3(P_1+r)^2}\,dx\,.\nonumber\\ \label{lambdaiii}
\eeqa
Equation (\ref{cambio}) maps the branches as follows: 
$y_1\leftrightarrow y_3$ and 
$y_2\leftrightarrow y_2$. Using
$y_3=\La_1^{\beta_1}\La_2^{\beta_2}/y_1$, we may 
express the expansion  for $dS_{SW}^{(3)}$ in terms of a comparable
one for $dS_{SW}^{(1)}$, for which $SU(N_1)\leftrightarrow SU(N_2)$, 
with the approximation (\ref{aprox}) exhibiting the branch cuts which connect 
branches 2 and 3.

Given the SW differential to the required accuracy,
we are able to compute the order parameters and 
dual order parameters to the appropriate order on both quantum scales, 
as they are 
given in terms of period integrals of the Seiberg--Witten differential. 
Nevertheless, as we pointed out earlier, we will avoid the computation
of the dual periods using the derivatives of the prepotential.

\section{Derivatives of the prepotential}

In $SU(N)$ gauge theories with matter in the fundamental
representation for which the 
Seiberg--Witten curve is a hyperelliptic one, the computation of the 
logarithmic derivative of the prepotential with respect to the quantum scale 
in terms of parameters of the curve \cite{hoker2,matone,eguchi,sonn} 
is found to be very useful to calculate
the instanton corrections to the prepotential \cite{hoker2,chan}. The reason
is that one does not need to compute the dual periods to obtain them
so the calculation is much simpler. In this section we will 
find a renormalization group type equation
that will help us to calculate the instanton corrections to the
prepotential. 

We will do this with the help of the Riemann bilinear relation, 
along the lines of what was done in $SU(N)$ gauge theories
\cite{eguchi,sonn}. The first thing we have to consider is 
that the effective 
(field dependent, dimensionless) gauge coupling
is given by the second derivative of the prepotential.
$\F$ is thus a homogeneous function of weight two on the variables 
$A_i=\{a_1,\,\cdots,\,a_{N_1},\,\hat a_1,\,\cdots,\,\hat a_{N_2}\},\,
M_j=\{m_1,\,\cdots,\,m_{N_{f_1}},\,\hat m_1,\,\cdots,\,\hat 
m_{N_{f_2}}\},$ and on the quantum scales of both groups 
$\La_1,\,\La_2$. Therefore satisfies the Euler equation
\be\label{ordendos}
2\F=\left(\La_1\frac{\pa}{\pa{\La_1}}+\La_2\frac{\pa}{\pa{\La_2}}+
\sum_{i=1}^{N_1+N_2} A_i\frac{\pa}{\pa{A_i}}
+\sum_{j=1}^{N_{f_1}+N_{f_2}} M_j\frac{\pa}{\pa{M_j}}\right)\F\,.
\ee
Taking the derivatives with respect to the moduli of the
curve $u_k$, and also using the definition of 
$A_D^i=\{a_{D}^1,\,\cdots,\,a_{D}^{N_1},\,\hat a_{D}^1,\,\cdots,\,\hat
a_D^{N_2}\}=\frac{\pa {\cal F}}{\pa A_i}$ one obtains
\be\label{der}
{\pa\over \pa{u_k}}\left(\La_1\frac{\pa}{\pa{\La_1}}+\La_2\frac{\pa}
{\pa{\La_2}}+\sum_j M_j\frac{\pa}{\pa{M_j}}
\right)\F=\sum_i\left(A_D^i{\pa\over\pa u_k}A_{i}-A_{i}{\pa\over\pa 
u_k}A_D^i\right)\,.
\ee

Using now the definitions of $A_i$, $A_D^i$ as 
the periods of the Seiberg--Witten differential we arrive at
\be\label{per}
{\pa\over \pa{u_k}}\left(\La_1\frac{\pa}{\pa{\La_1}}+\La_2 \frac{\pa}
{\pa\La_2}+\sum_j M_j\frac{\pa}{\pa{M_j}}
\right)\F
=\sum_{i=1}^{N_1+N_2}\oint_{\alpha_i}d\omega_k\,\oint_{\beta^i}dS_{SW}-
\oint_{\beta^i}d\omega_k\,\oint_{\alpha_i}dS_{SW}\,,
\ee
where we have used $\pa_{u_k}dS_{SW}=d\omega_k$.

The right hand side of this equation can be evaluated with the help of
a Riemann bilinear relation \cite{bilinear}. In a general case, 
if we have two abelian
differentials $d\Omega$ and $d\Omega'$ defined on a genus $g$ Riemann 
surface, the Riemann bilinear relations read
\be\label{bil}
\sum_{i=1}^g\,\oint_{\alpha_i}d\Omega\,\oint_{\beta^i}d\Omega'-
\oint_{\beta^i}d\Omega\,\oint_{\alpha_i}d\Omega'=
\frac{1}{2\pi i}\sum_{s_\lambda}{\rm res}_{s_\lambda} (\Omega d\Omega')
\ee
where, when $d\Omega$ is a holomorphic differential, $s_\lambda$ denote 
the poles of $d\Omega'$. In this case 
we have $d\Omega=d\omega_k$ and $d\Omega'=  
dS_{SW}$. As can be seen from (\ref{ds}) the Seiberg--Witten differential 
$dS_{SW}$ is an abelian differential of the third kind with 
 poles at $x \rightarrow\infty$ and $x \rightarrow m_j,\,\hat m_j$, 
and also $d\omega_k$ is the holomorphic differential (\ref{omega}). 
Then $s_\lambda=\{\infty,m_j,\hat m_j\}$. 
As our Riemann surface is a three branched cover of the sphere, 
we have that
\be
d\omega_k\stackrel{x\rightarrow\infty}\longrightarrow\left\{
\begin{array}{l}
\,\,x^{-k}dx+{\cal O}(x^{-k-1})
\hspace{1.67cm}\mbox{branch\,1}\\
-x^{-k}dx+{\cal O}(x^{-k-1})
\hspace{1.5cm}\mbox{branch\,2}\\
\,\,\,{\cal O}(x^{-\beta_2-k})
\hspace{2.96cm}\mbox{branch\,3}
\end{array}\right.\,,\label{uno}
\ee
and
\be
dS_{SW}\stackrel{x\rightarrow\infty}\longrightarrow\left\{
\begin{array}{l}
\,\,(N_1-N_{f_1})dx+{\cal O}(x^{-1})\hspace{1.07cm}\mbox{branch 1}\\
-{(N_1-N_2)}dx+{\cal O}(x^{-1})\hspace{1.05cm}\mbox{branch 2}\\
\,\,\,{\cal O}(1)\hspace{4.2cm}\mbox{branch 3}
\end{array}\right.\,.\label{dos}
\ee

Also when $x\rightarrow m_j,\,\hat m_j$
\be
dS_{SW}\stackrel{x\rightarrow m_j}\longrightarrow\left\{
\begin{array}{l}
\,\,-{m_j\over x-m_j}dx+\cdots\\
\,\,\,\cdots\\
\,\,\,\cdots
\end{array}\right.\hspace{.3cm},\hspace{.8cm}
dS_{SW}\stackrel{x\rightarrow \hat m_j}\longrightarrow\left\{
\begin{array}{l}
\,\,\,\cdots \hspace{3.0cm}\mbox{branch\,1}\\
{\hat m_j\over x-\hat m_j}dx+\cdots
\hspace{1.4cm}\mbox{branch\,2}\\
\,\,\,\cdots
\hspace{3.0cm}\mbox{branch\,3}
\end{array}\right.
\,,\label{tres}
\ee
where by $\cdots$ we denote terms that are regular when
$x\rightarrow m_j,\,\hat m_j$. Now, using (\ref{bil})--(\ref{tres}) 
 we can write Eq.(\ref{per}) as
\beqa\label{peri}
&&{\pa\over \pa{u_k}}\left(\La_1\frac{\pa}{\pa{\La_1}}+\La_2 \frac{\pa}
{\pa\La_2}+\sum_j M_j\frac{\pa}{\pa{M_j}}\right)\F
={\beta_1\over2\pi i}\delta_{k,2}-{1\over2\pi i}
\sum_jM_j\,\omega_k\mid_{x=M_j}+\nonumber\\
&&+{1\over2\pi i}\sum_jM_j\,\omega_k\mid_{x=\infty}
={\beta_1\over2\pi i}\delta_{k,2}+{1\over2\pi i}
\sum_jM_j\int_{M_j}^\infty d\omega_k\,.
\eeqa

We now want to further simplify this formula. Using $\frac{\pa \F}
{\pa u_k}=\sum_{i}\frac{\pa A_i}{\pa u_k}A_D^i$ we get to
\be
\sum_j M_j\frac{\pa \F}{\pa{M_j}\pa u_k}=\sum_{i,j}M_j\frac{\pa A_i}
{\pa u_k}\frac{\pa A_D^i}{\pa M_j}
=\sum_{i,j}M_j\oint_{\alpha_i}d\omega_k\oint_{\beta^i} \frac{\pa
 dS_{SW}}{\pa M_j}\,,
\ee
so now we can use again the Riemann bilinear relation and we arrive to
\be
\sum_{i,j}M_j\oint_{\alpha_i}d\omega_k\oint_{\beta^i} \frac{\pa
 dS_{SW}}{\pa M_j}=\sum_jM_j\sum_{s_\lambda}
{\rm res}_{s_\lambda} \omega_k \frac{\pa dS_{SW}}{\pa M_j}\,,
\ee
as $\oint_{\alpha_i}\frac{\pa dS_{SW}}{\pa M_j}=\frac{\pa A_i}{\pa
 M_j} =0$. Furthermore, as can be read off from (\ref{dos}), (\ref{tres}),
 the differential $\frac{\pa dS_{SW}}{\pa M_j}$ is a third kind
 differential with a pole of order $1$ in $x=\infty,\,M_j$ with 
residue $+1$ and $-1$ respectively. Therefore
\be
\sum_{s_\lambda}
{\rm res}_{s_\lambda} \omega_k \frac{\pa dS_{SW}}{\pa M_j}=
\omega_k\mid_{\infty}-\omega_k\mid_{M_j}=\int_{M_j}^\infty d\omega_k\,.
\ee
So we finally arrive to the expression
\be
{\pa\over \pa{u_k}}\left(\La_1\frac{\pa}{\pa{\La_1}}+\La_2 \frac{\pa}
{\pa\La_2}\right)\F={\beta_1\over2\pi i}\delta_{k,2}\label{del}\,.
\ee

Then, integrating Eq.(\ref{del}) with respect to $u_k$ we get
\be\label{betauno}
\left(\frac{\pa}{\pa{\log\La_1}}+\frac{\pa}{\pa{\log\La_2}}
\right)\F={\beta_1\over2\pi i} u_2\, +\, 
\mbox{terms indep. of}\,\,u_k\,.
\ee

We can actually perform a similar calculation but taking the derivative 
of (\ref{ordendos}) with respect to the moduli $\hat u_k$, 
and we will get to
\be\label{betados}
\left(\frac{\pa}{\pa{\log\La_1}}+\frac{\pa}{\pa{\log\La_2}}
\right)\F={\beta_2\over2\pi i} \hat u_2\, +\, 
\mbox{terms indep. of}\,\hat u_k\,.
\ee

Therefore, from (\ref{betauno}) and (\ref{betados}), 
we can conclude that
\beqa
&&\left(\frac{\pa}{\pa{\log\La_1}}+\frac{\pa}{\pa{\log\La_2}}
\right)\F=
{1\over2\pi i}\left(\beta_1 u_2+\beta_2 \hat u_2 \right)\, 
+\, \mbox{terms indep. of}\,\,u_k \,,\hat u_k=\nn\\
&&\hspace{1cm}={1\over2\pi i}\left(\frac{\beta_1}{2} \sum_{i=1}^{N_1}
e_i^2+\frac{\beta_2}{2} \sum_{i=1}^{N_2}\hat e_i^2 \right)\, 
+\, \mbox{terms indep. of}\,\,e_k \,,\hat e_k\,,
\label{beta}
\eeqa
where we have used $2u_2= \sum_{i=1}^{N_1}e_i^2$, $2\hat u_2=
 \sum_{i=1}^{N_2}\hat e_i^2$. The terms independent of $u_k$, $\hat
u_k$, are unphysical constant terms that depend on the exact
parametrization of the prepotential.

It is interesting to point out that the final result appears to 
be just the sum of the equations for $SU(N_1)$ and $SU(N_2)$. 
Nevertheless that is not the case, as $u_2$ depends both
on $a_i$ and $\hat a_i$, and the same happen for $\hat u_2$. That 
means that there
is a non-trivial mixing between both groups, as is expected in the
presence of a hypermultiplet in the bifundamental representation.

\section{Instanton corrections to the prepotential}

The quantum relations between the low-energy coordinates of the moduli
space $a_i,\,\hat a_i$, $a_D^j,\,\hat a_D^j$, and the 
parameters on the curve,
are implicitly given by the period integrals
\be
a_k=\oint_{\alpha_k}dS_{SW}^{(1)} \,\,,\hspace{1.5cm}
\hat{a}_k=\oint_{\alpha_{N_1+k}}dS_{SW}^{(3)} \label{a}\,,
\ee
and \be
a_D^k=\oint_{\beta^k}dS_{SW}^{(1)} \,\,,\hspace{1.5cm}
\hat{a}_D^k=\oint_{\beta^{N_1+k}}dS_{SW}^{(3)}\,, \label{ad}
\ee
where $(\alpha_i,\beta^j)$ constitute a symplectic basis of homology
cycles of the Riemann surface with canonical intersections. 
The homology cycles $\alpha_k$ and $\beta_k$, 
$k=1,\cdots,N_1$, are defined for Riemann branches $y_1$ and $y_2$, 
and cycles ${\alpha}_k$ and ${\beta}_k$, $k=N_1+1,\cdots,N_1+N_2$, 
for Riemann branches $y_2$ and $y_3$. 
The cycle $\alpha_k$ is chosen to be a simple contour 
enclosing the branch cut connecting $e_k^+$ with $e_k^-$ 
($k=1,\cdots,N_1$) on 
branch 1, while for $k=1,\cdots,N_2$ similarly encloses 
the branch cut connecting $\hat{e}_k^+$ with $\hat{e}_k^-$ 
on branch 3. The
canonical prepotential ${\cal F}$ is then implicitly defined by the
equation
\be
a_D^i(a_j,\hat a_k)=\frac{\partial {\cal F}}{\pa a_i}\,\,,\hspace{1.5cm}
\hat a_D^i(a_j,\hat a_k)=\frac{\partial {\cal F}}{\pa \hat a_i}\,,
\ee
so that its exact determination involves the integration of the functions
$a_D^i(a_j,\hat a_k)$, $\hat a_D^i(a_j,\hat a_k)$ for which there is no
closed form available. In this context the existence of an algorithm
that enables us to determine the exact form of ${\cal F}$ without going
through the actual computation of the dual periods is welcome.

The holomorphic prepotential can be expressed as 
\be\label{efe}
{\cal F}={\cal F}_{classic}+{\cal F}_{1-loop}+{\cal F}_{instanton}\,,
\ee
since the perturbative corrections saturate at one--loop in ${\cal N}=2$
theories \cite{Sei} but there is an infinite series of 
non-perturbative instanton contributions. The one-loop 
perturbative correction to the prepotential 
has been calculated in \cite{sch3} for the case of product gauge groups 
by calculating the period integrals using the hyperelliptic 
approximation, and is given by
\beqa
{\cal F}_{1-loop}&=&{i\over 4\pi}\left\{\sum_{{i,j=1}\atop{i<j}}^{N_1}\,
(a_i-a_j)^2\,{\rm log}\,(a_i-a_j)^2+
\sum_{{\alpha,\beta=1}\atop{\alpha<\beta}}^{N_2}\,
(\hat a_{\alpha}-\hat a_{\beta})^2\,{\rm log}\,(\hat a_{\alpha}-
\hat a_{\beta})^2\,\right.\nonumber\\
&-&{1\over 2}\sum_{i=1}^{N_1}\,\sum_{\alpha=1}^{N_2}\,
(a_i-\hat a_{\alpha})^2\,{\rm log}\,(a_i-\hat a_{\alpha})^2-
{1\over 2}\sum_{i=1}^{N_1}\sum_{f=1}^{N_{f_1}}\,(a_i-m_f)^2
\,{\rm log}\,(a_i-m_f)^2\,-\nonumber\\
&-&\left.{1\over 2}\sum_{\alpha=1}^{N_2}\sum_{f=1}^{N_{f_2}}\,
(\hat a_{\alpha}-\hat m_f)^2\,{\rm log}\,
({\hat a_{\alpha}-\hat m_f)^2} \right\}\,. \label{oneloop}
\eeqa

For the non-perturbative corrections in (\ref{efe}), note that the 
n-th order instanton corrections to the prepotential
 is not just the sum of contributions proportional to $\La_1^{n\beta_1}$, 
$\La_2^{n\beta_2}$, but all possible combinations of terms $\La_1^{n_1\beta_1}
\La_2^{n_2\beta_2}$ with $n_1+n_2=n$. Therefore we will take the form
of ${\cal F}_{instanton}$ to be 
\be
{\cal F}_{instanton}(a_i,\hat a_i)=\frac{1}{2\pi i}
\sum_{k,l=1}^\infty {\cal F}_{k,l}
(a_i,\hat a_i)\La_1^{k\beta_1}\La_2^{l\beta_2}\,.\label{instan}
\ee
We will check this assumption in the next subsection.

\subsection{Computation of instanton corrections}

To compute the instanton corrections to the prepotential we 
need to perform the period integral (\ref{a}) for the order parameters. 
Those periods can be computed 
by reducing the evaluation of the integrals to a set of residue
calculations in the same way as it was first done in \cite{hoker1}. 
The argument is the following: We can always 
consider the distance between the contour $\alpha_k$ and the
 branch cut connecting $e_k^+\,(\hat e_k^+)$ with $e_k^-\,(\hat e_k^-)$ 
to be much larger than $\La_1^{\beta_1}$, $\La_2^{\beta_2}$. Then, 
if we expand $dS_{SW}$ in 
power series around $\La_1=0$ and $\La_2=0$, the integrals (\ref{a})
over the cycle $\alpha_k$ can be performed just by calculating 
residues at the points 
$e_k$ (for $\alpha_k$, $k=1,\cdots,N_1$), or $\hat e_k$ (for 
$\alpha_k$, $k=N_1+1,\cdots,N_1+N_2$).

Therefore, if we want to compute in this way the order 
parameters $a_i$ we need to expand
$dS_{SW}^{(1)}$ around $\La_1=0$, $\La_2=0$. Using (\ref{lambdai})--
(\ref{lambdaiii}) we get
\beqa
\left(dS_{SW}^{(1)}\right)_I&\simeq& x\frac{\pa_x P_1}{P_1}+
\sum_{k=1}^\infty \frac{(2k-1)!}{(k!)^2}\left(\frac{P_0P_2}{P_1^2}
\right)^k\La_1^{k\beta_1}\label{dsuno}\,,\\
\left(dS_{SW}^{(1)}\right)_{II}&\simeq& -P_3\sum_{k=1}^\infty
\frac{(2k)!}{(k-1)!(k+1)!}\frac{P_0^{k+1}P_2^{k-1}}{P_1^{2k+1}}
\La_1^{(k+1)\beta_1}\La_2^{\beta_2}\label{dsdos}\,,\\
\vdots && \hspace{2cm} \nonumber\vdots
\eeqa
where by $\simeq$ we denote terms
up to total derivatives or terms that have no residue at the
corresponding cycles.

Using this we can write the order parameters $a_i$ as
\be
a_i= {\rm res}_{e_i}dS_{SW}^{(1)}=
e_i+\sum_{k=1}^{\infty}\Delta^{(k)}_i(e_i)\La_1^{\beta_1k}
+ \sum_{k,l=1}^{\infty}\Delta^{(k,l)}_i(e_i)\La_1^{\beta_1(k+2l-1)}
\La_2^{\beta_2l}\,,\label{expuno}
\ee
where the exact form of the functions ${\Delta}^{(k)}_i$, $\hat
{\Delta}^{(k)}_i$, $\Delta^{(k,l)}_i$, $\hat\Delta^{(k,l)}_i$ is 
obtained using (\ref{dsuno}), (\ref{dsdos}), and is given by
\beqa
\Delta^{(k)}_i(x)&=&\frac{1}{(k!)^2} 
\left({\pa\over{\pa x}}\right)^{2k-1} \bigl(\frac{P_0(x)^kP_2(x)^k}
{\prod_{j \neq i}(x-e_j)^{2k}}\bigr)\,,\\
\Delta^{(k,1)}_i(x)&=&-\frac{k}{k!(k+1)!}\left(\frac{\pa}{\pa x}\right)^{2k}
\bigl(P_3(x)\frac{(P_0(x))^{k+1} (P_2(x))^{k-1}}
{\prod_{j \neq i}(x-e_j)^{2k+1}}\bigr)\,,\label{pr}\\
\vdots &\,& \hspace{2cm} \nonumber\vdots
\eeqa

Also by a similar computation we obtain
\be 
\hat a_i= \hat e_i+\sum_{k=1}^{\infty}\hat \Delta^{(k)}_i(\hat e_i)
\La_2^{\beta_2k}+
\sum_{k,l=1}^{\infty}\hat\Delta^{(k,l)}_i(\hat e_i)\La_1^{\beta_1k}
\La_2^{\beta_2(l+2k-1)}\,,\label{expa}
\ee
where
\beqa
\hat\Delta^{(k)}_i(x)&=&\frac{1}{(k!)^2} 
\left({\pa\over{\pa x}}\right)^{2k-1} \bigl(\frac{P_1(x)^kP_3(x)^k}
{\prod_{j \neq i}(x-\hat e_j)^{2k}}\bigr)\,,\\
\hat\Delta^{(k,1)}_i(x)&=&-\frac{k}{k!(k+1)!}\left(\frac{\pa}
{\pa x}\right)^{2k}
\bigr(P_0(x)\frac{(P_3(x))^{k+1} (P_1(x))^{k-1}}
{\prod_{j \neq i}(x-\hat e_j)^{2k+1}}\bigl)\,,\label{prdos}\\
\vdots &\,& \hspace{2cm} \nonumber\vdots
\eeqa

Once we have the expression for the power expansion of the order 
parameters in $\La_1$, $\La_2$ one can use the 
equation (\ref{beta}) to compute the instanton correction to the
prepotential. Just by inserting the expression for the 
prepotential (\ref{efe}) in Eq.(\ref{beta}) one gets
\be
\frac{\beta_1}{2}\sum_ia_i^2+\frac{\beta_2}{2}\sum_i\hat a_i^2 +
\sum_{k,l=1}^\infty (\beta_1k+\beta_2l){\cal F}_{k,l}(a_j,\hat a_j)
\La_1^{\beta_1k}\La_2^{\beta_2l}=\frac{\beta_1}{2}\sum_ie_i^2+
\frac{\beta_2}{2}\sum_i\hat e_i^2\,\,,\label{exdos}
\ee
up to unphysical constant terms.

From here it is clear how to extract the instanton corrections to
the prepotential from (\ref{exdos}). The first thing we need is 
the expansions (\ref{expuno}),
(\ref{expa}) of the order parameters in terms of $e_i$, $\hat
e_i$. Secondly, let us expand ${\cal F}_{k,l}(a_j,\hat a_j)$ in power series
around $a_i=e_i$, and $\hat a_i=\hat e_i$. We have
\beqa 
&&{\cal F}_{k,l}(a_j,\hat a_j)={\cal F}_{k,l}(e_j,\hat e_j)+
\sum_i\pa_{e_i}{\cal F}_{k,l}\Delta^{(1)}_i\La_1^{\beta_1}+
\sum_i\pa_{\hat e_i}{\cal F}_{k,l}\hat\Delta^{(1)}_i
\La_2^{\beta_2}+(\sum_i\pa_{e_i}{\cal F}_{k,l}\Delta^{(2)}_i+\nonumber\\
&&+\frac12\sum_{i_1,i_2}\pa_{e_{i_1}}\pa_{e_{i_2}}{\cal F}_{k,l}
\Delta^{(1)}_{i_1}\Delta^{(1)}_{i_2})\La_1^{2\beta_1}+
(\sum_i\pa_{\hat e_i}{\cal F}_{k,l}\hat\Delta^{(2)}_i+
\frac12\sum_{i_1,i_2}\pa_{\hat e_{i_1}}\pa_{\hat e_{i_2}}{\cal F}_{k,l}
\hat\Delta^{(1)}_{i_1}\hat\Delta^{(1)}_{i_2})\La_2^{2\beta_2}+\nonumber\\
&&+\sum_{i_1,i_2}\pa_{e_{i_1}}\pa_{\hat e_{i_2}}{\cal F}_{k,l}
\Delta^{(1)}_{i_1}\hat\Delta^{(1)}_{i_2}\La_1^{\beta_1}\La_2^{\beta_2}+\cdots
\label{efekl}\eeqa

Now inserting (\ref{expuno}), (\ref{expa}) and (\ref{efekl}) into 
(\ref{exdos}) it is possible to obtain the instanton corrections 
recursively just by identifying order by order in $\La_1$, $\La_2$ in
Eq.(\ref{exdos}). For example for the first
corrections we have~\footnote{We just list one half of the instanton
corrections here. The other half are obtained from these ones just by
using the fact that the flip $SU(N_1)\leftrightarrow SU(N_2)$
interchanges ${\cal F}_{k,l}\leftrightarrow{\cal F}_{l,k}$. Note that we only 
get contributions of the form (\ref{instan}) to the instanton corrections.}
\beqa
&&{\cal F}_{1,0}(a_j,\hat a_j)=-\sum_{i=1}^{N_1}a_i\Delta^{(1)}_i\,\,,\,\,\,
\,\,{\cal F}_{2,0}(a_j,\hat a_j)=-\frac12\sum_{i=1}^{N_1}(a_i\Delta^{(2)}_i+
\frac12\Delta^{(1)}_i\Delta^{(1)}_i+
\pa_{a_i} {\cal F}_{1,0}\Delta^{(1)}_i)\,,\nonumber\\
&&{\cal F}_{1,1}(a_j,\hat a_j)=-
\frac{\beta_2}{\beta_1+\beta_2}\sum_{i=1}^{N_1}\pa_{a_i}{\cal F}_{0,1}
\Delta^{(1)}_i-\frac{\beta_1}{\beta_1+\beta_2}
\sum_{i=1}^{N_2}\pa_{\hat a_i} {\cal F}_{1,0}\hat\Delta^{(1)}_i
\,,\nonumber\\
&&{\cal F}_{3,0}(a_j,\hat a_j)=-\frac13\sum_{i=1}^{N_1}(a_i\Delta^{(3)}_i+
\Delta^{(1)}_i\Delta^{(2)}_i+2\pa_{a_i} {\cal F}_{2,0}\Delta^{(1)}_i+
\pa_{a_i} {\cal F}_{1,0}\Delta^{(2)}_i+\nonumber\\
&&\hspace{2cm}+\sum_{j=1}^{N_1}
\frac12\pa_{a_i}\pa_{a_j} {\cal F}_{1,0}\Delta^{(1)}_i\Delta^{(1)}_j)\,,
\nonumber\\
&&{\cal F}_{2,1}(a_j,\hat a_j)=-\frac{\beta_1}{2\beta_1+\beta_2} 
\sum_{i=1}^{N_1}(a_i\Delta^{(1,1)}_i+2\pa_{\hat a_i} {\cal F}_{2,0}\hat
\Delta^{(1)}_i+\sum_{j=1}^{N_2}\pa_{a_i}\pa_{\hat a_j}{\cal F}_{1,0}
\Delta^{(1)}_i\hat\Delta^{(1)}_j)-\nonumber\\
&&\hspace{.8cm}-\frac{\beta_2}{2\beta_1+\beta_2}\sum_{i=1}^{N_1}(\pa_{a_i}
{\cal F}_{0,1}\Delta^{(2)}_i+\frac12\sum_{j=1}^{N_2}\pa_{a_i}\pa_{a_j}
{\cal F}_{0,1}\Delta^{(1)}_i\Delta^{(1)}_j)-\frac{\beta_1+\beta_2}
{2\beta_1+\beta_2}\sum_{i=1}^{N_1}
\pa_{a_i}{\cal F}_{1,1}\Delta^{(1)}_i\nonumber\,,\\
&&\hspace{1cm}\vdots\hspace{6cm}\vdots\label{cosas}
\eeqa

Note that the method developed here allow us to compute the
instanton corrections to the prepotential recursively in a remarkably 
straightforward way up to arbitrary high orders. In fact, 
Eq.(\ref{exdos}) together with the hyperelliptic approximation $(dS_{SW})_I$
to the Seiberg--Witten differential, fixes the instanton corrections up to 
order two in both quantum scales ({\it i.e.}up to terms $\La_1^{k\beta_1}
\La_2^{l\beta_2}$ with $k+l=2$). Also the next correction to the Seiberg--
Witten differential $(dS_{SW})_{II}$ is enough to reach the fifth order 
instanton correction. 

\subsection{Examples}

In this subsection we will write explicitly the instanton
corrections obtained for some product gauge groups.

\vspace{.5cm}

\noindent{\bf \underline{$SU(2)\times SU(2)$}}

\vspace{.3cm}

For asymptotically free theories, the most general case that one can
consider is the case with one 
matter hypermultiplet in the fundamental representation of each 
group and one in 
the bifundamental. We obtain the following results
\beqa
&&{\cal F}_{1,0}=-\frac{m\hat u}{2u}\,\,,\,\,\,\,\,\,\,\,{\cal F}_{0,1}=
-\frac{\hat m u}{2\hat u}\,\,,\,\,\,\,\,\,\,
{\cal F}_{1,1}=\frac{m\hat m(u+\hat u)}{4u\hat u}\nn\,\,,\\
&&{\cal F}_{2,0}=\frac{2u^2\hat u -3u\hat u^2+m^2(u^2+ 5\hat
u^2-6u\hat u)}{64u^3}\,,\,\,
{\cal F}_{0,2}=\frac{2u\hat u^2 -3u^2\hat u+\hat m^2(\hat u^2+ 5
u^2-6u\hat u)}{64\hat u^3}\,,\nonumber\\
&&{\cal F}_{2,1}=-\frac{\hat m(u-\hat u)(u^2-5\hat um^2+u(3\hat
u+m^2))}{64u^3 \hat u}\,\,,\\
&&\F_{2,2}=\frac{u^4(5\hat m^2-3\hat u)+u^3(11\hat u^2+25m^2\hat
m^2-3\hat u(5m^2+11\hat m^2))+u^2(11\hat u^3+15\hat um^2\hat m^2)}
{1024u^3\hat u^3}\nn\\
&&+\frac{-5u^2\hat u^2(m^2+\hat m^2)-3u(\hat u^4-5\hat u^2m^2\hat m^2+\hat
u^3(11m^2+5\hat m^2))+5\hat u^3m^2(\hat u+5\hat m^2)}{1024u^3\hat u^3}\,,
\label{che}\eeqa
where $u=-a_1a_2=a_1^2$ and $\hat u=-\hat a_1\hat a_2=\hat a_1^2$.

From these results it is possible to extract the results for the case
with less matter hypermultiplets in the fundamental. This result is
obtained just by taking $m$ ($\hat m$) to infinity while keeping
$m\La_1$ ($\hat m\La_2$) fixed to the new scale. For example we have that
for $SU(2)\times SU(2)$ without matter in the fundamental one gets
\beqa
&&{\cal F}_{1,0}=-\frac{(u+ \hat 
u)}{2u}\,,\,\,\,\,\,\,\,\,\,\,{\cal F}_{0,1}=
-\frac{(u+ \hat u)}{2\hat u}\,,\,\,\,\,\,\,\,\,\,
{\cal F}_{1,1}=\frac{(u+\hat u)}{4u\hat u}\nonumber\\
&&{\cal F}_{2,0}=\frac{u^2+ 5\hat u^2-6u\hat u}{64u^3}\,,\,\,\,\,\,
{\cal F}_{0,2}=\frac{\hat u^2+ 5 u^2-6u\hat u}{64\hat u^3}\,\,,\nonumber\\
&&{\cal F}_{2,1}=-\frac{ u^2+ 5 \hat u^2-6u\hat u}{64u^3 \hat u}
\label{chec}\nn\,\,,
\eeqa

\vspace{.3cm}

\noindent{\bf\underline{$SU(3)\times SU(2)$}}

\vspace{.3cm}

In this example, for asymptotically free theories the most general 
case that one can consider is the case with two 
matter hypermultiplets in the fundamental of $SU(3)$. 
We obtain the following results
\beqa
{\cal F}_{1,0}&=&\frac{-2u^3+2m_1m_2u^2-3(m_1+m_2)uv+18v^2-2u^2\hat u
-6m_1m_2u\hat u+9(m_1+m_2)v\hat u}{4u^3-27v^2}\,,\nonumber\\
{\cal F}_{0,1}&=&-\frac{v}{2\hat u}\,,\,\,\,\,\,\,\,\,
{\cal F}_{2,0}=\frac{5v^2+2u\hat u^2-3\hat u u^2+5v^2}{64\hat u^3}\,,
\nonumber\\
{\cal F}_{1,1}&=&\frac{2\hat u(2u^2(m_1+m_2)-3uv)+2vu^2+6v(u-3\hat u)
m_1m_2-9v^2(m_1+m_2)}{2\hat u(4u^3-27v^2)}\,,
\eeqa
where $u=-a_1a_2-a_1a_3-a_2a_3$, $v=a_1a_2a_3$ and $\hat u=-\hat
a_1\hat a_2= \hat a_1^2$.

As in the previous case, from these calculations it is possible to extract 
the results for the case with less matter in the fundamental. This result is
obtained by taking $m_1,m_2$ to infinity while keeping
$m_{1,2}\La_1$ fixed to the new scale.

\subsection{Some checks}

In this subsection we will perform some non-trivial checks to the
results obtained previously. The first check we can perform is to
compare our results with the ones presented in \cite{sch3} for the
first instanton correction to the prepotential. 
Note that the first instanton correction obtained by us in 
(\ref{cosas}) can be rewritten in the following way\footnote{Using 
\be
\sum_{i=1}^{N_1}{\rm res}_{e_i}x\frac{P_0(x)P_2(x)}{(P_1(x))^2}=
\sum_{i=1}^{N_1}e_i\Delta_i^{(k)}(e_i)+\sum_{i=1}^{N_1}
\frac{P_0(e_i)P_2(e_i)}{\prod_{k\neq i}(e_i-e_k)^2}=0\,,\label{res}
\ee
up to constant unphysical terms that come from the residue at infinity of the
function $x\frac{P_0(x)P_2(x)}{(P_1(x))^2}$.}. 
\be
{\cal F}_{1,0}=-\sum_{i=1}^{N_1}a_i\Delta_i^{(k)}=\sum_{i=1}^{N_1}
\frac{P_0(a_i)P_2(a_i)}{\prod_{k\neq i}(a_i-a_k)^2}\,,
\ee
and an analogous expression for ${\cal F}_{0,1}$. Then, once we 
take $\hat m_k=0$ ($m_k=0$), we get the 
same result for the first instanton correction as the one in
\cite{sch3} for an $SU(N_1)\times SU(N_2)$ with massless
hypermultiplets in the fundamental of both groups.

For higher instanton corrections the only result available in the
literature is the one in \cite{feich} for
the case $SU(2)\times SU(2)$. They compute up to the third 
instanton correction using the fact that for this particular case the
curve is still hyperelliptic, so one can use the standard
Picard--Fuchs techniques
to calculate the instanton corrections. We have checked that 
our results for that case listed in
(\ref{chec}) agree with those presented in \cite{feich}.

Also, as we explained in the previous sections, another check that we can
perform is that in the
limit $\La_2\rightarrow 0$ one should recover the instanton
corrections for a
${\cal N}=2$ supersymmetric Yang--Mills theory with gauge group
$SU(N_1)$ and $N_{f_1}+N_2$ matter hypermultiplets in the fundamental
representation of the gauge group. In fact in that case we have
\beqa
a_i&=& e_i + \sum_{k=1}^{\infty} \left({\pa\over{\pa e_i}}\right)^{2k-1}  
\Delta_i^{(k)}(e_i)\La_1^{\beta_1}\,\,,\\
\Delta_i^{(k)}(x)&=&\frac{1}{(k!)^2}
(\frac{\prod_{f=1}^{N_{f_1}}(x+m_f)\prod_{l=1}^{N_2}(x+\hat e_l)}
{\prod_{j \neq k}(x-e_j)^2})^k\,,
\eeqa
where now the quantities $\hat a_i$ play the role of masses in the
fundamental of $SU(N_1)$, so they do not 
have $\Lambda_1$ corrections and they are just given by $\hat a_i=\hat
e_i$. From here we have
\beqa
&&{\cal F}_{1,0}=-\sum_{i=1}^{N_1}a_i\Delta^{(1)}_i\,,
\hspace{.5cm}
{\cal F}_{2,0}=-\frac12\sum_{i=1}^{N_1}(a_i\Delta^{(2)}_i+
\frac12\Delta^{(1)}_i\Delta^{(1)}_i+
\pa_{a_i} {\cal F}_{1,0}\Delta^{(1)}_i)\,,\nonumber\\
&&{\cal F}_{3,0}=-\frac13\sum_{i=1}^{N_1}(a_i\Delta^{(3)}_i+
\Delta^{(1)}_i\Delta^{(2)}_i+2\pa_{a_i} {\cal F}_{2,0}\Delta^{(1)}_i+
\pa_{a_i} {\cal F}_{1,0}\Delta^{(2)}_i+\nonumber\\
&&\hspace{2cm}+\sum_{j=1}^{N_1}
\frac12\pa_{a_i}\pa_{a_j} {\cal F}_{1,0}\Delta^{(1)}_i\Delta^{(1)}_j)\,,
\nonumber\\
&&\hspace{1cm}\vdots\hspace{6cm}\vdots\label{ccc}
\eeqa
that are exactly the instanton corrections for a
${\cal N}=2$ supersymmetric Yang--Mills theory with gauge group
$SU(N_1)$ and $N_{f_1}+N_2$ matter hypermultiplets in the fundamental
representation of the gauge group, as one would expect. Also the same
is true when $\La_2\rightarrow0$. This limit fixes completely
the first instanton correction to the prepotential. In fact, the result
(\ref{ccc}) agrees with the results available in the literature 
for these cases (see for example \cite{chan,mas1,mas2,hoker1}).

Another non-trivial check of the result that we can perform is 
that in the case $N_1=N_2=N$, in the limit 
$\La_1,\La_2\rightarrow 0$, $\La_1\La_2\rightarrow \La^2$ and
 $a_i=\hat a_i$ one should
recover the result for $SU(N)$ with $N_{f_1}+N_{f_2}$ matter
hypermultiplets in the fundamental representation. Taking this 
limit in (\ref{che}) we get 
\beqa
&&{\cal F}_{1,1}\longrightarrow\frac{m\hat m}{2u}\equiv{\cal
 F}_{1}^{SU(2)}\nonumber\,,\\
&&\F_{2,2}\longrightarrow\frac{u^2-3u(m^2+\hat m^2)+5m^2\hat
 m^2}{64u^3}\equiv{\cal F}_{2}^{SU(2)}\,,\nonumber
\eeqa
where $u=-a_1a_2=a_1^2$. Those are exactly the expected results.

\section{Generalization to higher representations of the gauge group}

In this section we will use the technique developed 
above to calculate the instanton corrections to the prepotential
for ${\cal N}=2$ super Yang--Mills theory with a matter 
hypermultiplet in the symmetric
and in the antisymmetric representation of the gauge group $SU(N)$.

\subsection{The symmetric representation}

The curve for this case is \cite{lopez2}
\be
y^3+P(x)y^2+x^2P(-x)\La^{N-2}+x^6\La^{3(N-2)}=0\,,
\ee
where $P(x)=\prod_{i=1}^N(x-e_i)$ denotes the characteristic
polynomial of $SU(N)$. Also we have $\sum e_i=\frac{N}{2} m$, 
$m$ being the mass of
the hypermultiplet in the symmetric representation. 
Therefore, 
this curve has the same form as the curve (\ref{curve}) just by 
identifying $P_0(x)=1$, 
$P_1(x)=P(x)$, $P_2(x)=x^2P(-x)$ and $P_3(x)=x^6$. Also one should 
take $\La_1=\La_2=\La$.

It is easy to see that, for this case Eq.(\ref{beta}) takes the form
\be
\frac12\sum_{i=1}^Na_i^2+\sum_{k=1}^\infty k{\cal F}_k(a_i)\La^{k(N-2)}
=\frac12\sum_{i=1}^Ne_i^2\,,
\ee
where
\beqa
a_i&=& e_i+\sum_{k=1}^{\infty}\Delta^{(k)}_i(e_i)\La^{k\beta}
+ \sum_{k,l=1}^{\infty}\Delta^{(k,l)}_i(e_i)\La^{\beta(k+3l-1)}\,,\\
\Delta^{(k)}_i(x)&=&\frac{1}{(k!)^2} 
\left({\pa\over{\pa x}}\right)^{2k-1} \bigl(\frac{x^{2k}P(-x)^k}
{\prod_{j \neq i}(x-e_j)^{2k}}\bigr)\,,\\
\Delta^{(k,1)}_i(x)&=&-\frac{k}{k!(k+1)!}\left(\frac{\pa}{\pa x}\right)^{2k}
\bigr(\frac{x^6(P(-x))^{k-1}}
{\prod_{j \neq i}(x-e_j)^{2k+1}}\bigl)\,,\\
&\vdots&\nonumber
\eeqa

Therefore the instanton corrections have the following the form
\beqa
&&{\cal F}_{1}(a_i)=-\sum_{i=1}^{N}a_i\Delta^{(1)}_i\,,
\hspace{.5cm}
{\cal F}_{2}(a_i)=-\frac12\sum_{i=1}^{N}\left(a_i\Delta^{(2)}_i+
\frac12\Delta^{(1)}_i\Delta^{(1)}_i+\pa_{a_i} {\cal F}_{1}\Delta^{(1)}_i
\right)\,,\nonumber\\
&&{\cal F}_{3}(a_i)=-\frac13\sum_{i=1}^{N}\bigl(a_i(\Delta^{(3)}_i+
\Delta_i^{(1,1)})+\Delta^{(1)}_i\Delta^{(2)}_i+
2\pa_{a_i} {\cal F}_{2}\Delta^{(1)}_i+
\pa_{a_i} {\cal F}_{1}\Delta^{(2)}_i+\nonumber\\
&&\hspace{2cm}+\sum_{j=1}^{N}\frac12\pa_{a_i}\pa_{a_j} 
{\cal F}_{1}\Delta^{(1)}_i\Delta^{(1)}_j\bigr)\,,\nonumber\\
&&\hspace{1cm}\vdots\hspace{6cm}\vdots
\eeqa
where to recover the mass dependence one should shift $a_i\rightarrow
a_i+\frac{m}{2}$. 

Note that by the same calculation as in (\ref{res}) we obtain 
\be
{\cal F}_{1}=-\sum_{i=1}^{N}a_i\Delta^{(1)}_i=
\sum_{i=1}^{N}\frac{a_i^2P(-a_i)}{\prod_{k \neq i}(a_i-a_k)^2}\,,
\ee
so we get the same result as in \cite{sch2} for the first instanton
correction. The first instanton correction for this case has been also
computed in \cite{slater} using the ADHM instanton calculus, and we
find also a pefect agreement with those results. 
There are no results available in the literature for
higher instanton corrections.

\subsection{The antisymmetric representation}

The curve for this case is \cite{lopez2}
\be
y^3+(x^2P(x)+3\La^{N+2})y^2+(x^2P(-x)+3\La^{N+2})\La^{N+2}+\La^{3(N+2)}
=0\,,
\ee
where $P(x)=\prod_{i=1}^N(x-e_i)$ denotes the characteristic
polynomial of $SU(N)$. Also we have $\sum e_i=-\frac{N}{2} m$, 
$m$ being the mass of
the hypermultiplet in the antisymmetric representation. 
Therefore, 
this curve has the same form as the curve (\ref{curve}) just by 
identifying $P_0(x)=1$, 
$P_1(x)=x^2P(x)+3\La^{N+2}$, $P_2(x)=x^2P(-x)+3\La^{N+2}$ 
and $P_3(x)=1$. Also one should take $\La_1=\La_2=\La$. This case is a
bit more complicated than the former ones because there is dependence
in $\La$ in $P_1$ and $P_2$ and therefore one has to be more careful in the
computation of the series expansion of $dS_{SW}$. In any case, one 
can still obtain the instanton corrections to the prepotential 
using the same procedure.

In fact, for the first terms of the series expansion of $a_i$ around
$\La=0$ we have
\be
a_i=e_i+\sum_{{m,n\geq0}\atop{m+n>0}}^N\Delta_i^{(m,n)}(e_i)\La^{\beta(m+n)}
+\cdots\,,\label{anti}
\ee
where
\be
\Delta_i^{(m,n)}(x)=\frac{(-3)^n}{(m!)^2n!}
\left(\frac{\pa}{\pa x}\right)^{2m+n-1}\frac{(x^2P(-x)+3\La^\beta)^m}
{x^{4m+2n}\prod_{k\neq i}(x-e_k)^{2m+n}}\La^{\beta(m+n)}\,,
\ee
being expression (\ref{anti}) exact up to order $\La^{3(N+2)}$.

Also here Eq.(\ref{beta}) takes the form
\be
\frac12\sum_{i=1}^Na_i^2+\sum_{k=1}^\infty k{\cal F}_k(a_i)\La^{k(N+2)}
=\frac12\sum_{i=1}^Ne_i^2\,,
\ee
from where we can extract the instanton corrections to the
prepotential. These are given by 
\beqa
&&{\cal F}_1=-\left.\sum_{i=1}^{N}a_i\pa_x\frac{P(-x)}
{x^{2}\prod_{k\neq i}(x-a_k)^{2}}\right|_{x=a_i}+\sum_{i=1}^{N}
\frac{3}{a_i\prod_{k\neq i}(a_i-a_k)}\,\,,\\
&&{\cal F}_2=-\left.
\frac12\sum_{i=1}^{N}\left(a_i(\frac{15}{2}\pa_x\frac{1}
{x^{4}\prod_{k\neq i}(x-a_k)^{2}}\right|_{x=a_i}-3\pa_x
\left.\frac{P(-x)}
{x^{4}\prod_{k\neq i}(x-a_k)^{3}}\right|_{x=a_i}+\right.\nn\\
&&+\frac{1}{4}\pa_x
\left.\frac{(P(-x))^2}
{x^{4}\prod_{k\neq i}(x-a_k)^{4}}\right|_{x=a_i})+
\frac12(\pa_x\left.\frac{P(-x)}
{x^{2}\prod_{k\neq i}(x-a_k)^{2}}\right|_{x=a_i}-\frac{3}{a_i^{2}
\prod_{k\neq i}(a_i-a_k)})^2+\nn\\
&&+\left.\left.
\pa_{a_i}{{\cal F}_{1}}(\pa_x\frac{P(-x)}
{x^{2}\prod_{k\neq i}(x-a_k)^{2}}\right|_{x=a_i}-\frac{3}{a_i^{2}
\prod_{k\neq i}(a_i-a_k)})
\right)\,\,,\\
&&\hspace{1cm}\vdots\hspace{6cm}\vdots\nn
\eeqa
where to recover the mass dependence one should shift $a_i\rightarrow
a_i-\frac{m}{2}$.

Note that now we have
\beqa
\sum_{i=1}^{N}{\rm res}_{e_i}x\frac{P(-x)}{(xP(x))^2}&=&
\sum_{i=1}^{N}\frac{P(-e_i)}{e_i^2\prod_{k \neq i}(e_i-e_k)^2}
+\left.\sum_{i=1}^{N}e_i\pa_x\frac{P(-x)}
{x^{2}\prod_{k\neq i}(x-e_k)^{2}}\right|_{x=e_i}
=-\frac{1}{P(0)}\,,\nn\\
\sum_{i=1}^{N}{\rm res}_{e_i}\frac{3}{xP(x)}&=&
\sum_{i=1}^{N}\frac{3}{e_i\prod_{k\neq i}(e_i-e_k)}=
-\frac{3}{P(0)}\,,\label{ansi}
\eeqa
where the last term in both equations in (\ref{ansi}) comes from the 
fact that the functions $x\frac{P(-x)}{(xP(x))^2}$ and $\frac{3}{xP(x)}$, 
have also a non--vanishing 
residue in $x=0$, not just in $x=e_i$. Therefore, we have
\be 
{\cal F}_1=\sum_{i=1}^{N}\frac{P(-a_i)}{a_i^2\prod_{k \neq
i}(a_i-a_k)^2}-\frac{2}{P(0)}\,.\label{aaa}
\ee
Note that this result agrees with the one presented in previous
calculations \cite{sch1}. There are no 
results available in the literature for higher instanton corrections.

We can also check this result if we take into account the fact 
that $SU(2)$ with matter
in the antisymmetric should give us the same result as pure
$SU(2)$. In fact what we find is
\be
{\cal F}_1=\frac{2}{u}\,\,,\hspace{1.5cm}{\cal F}_2=\frac{5}{4u^3}\,\,,
\ee
where $u=-a_1a_2=a_1^2$. 
This is exactly the result for pure $SU(2)$ obtained for example in
\cite{mas1,mas2} with a change in the quantum 
scale $\La^4\rightarrow\frac{\La^4}{4}$. Also for $SU(3)$ with one
hypermultiplet in the antisymmetric we should get the same result as
$SU(3)$ with one hypermultiplet in the fundamental. In fact we get
\be
{\cal F}_1=4\frac{6um-9v}{4u^3-27v^2}\,\,,
\ee
where $u=-a_1a_2-a_1a_3-a_2a_3$ and $v=a_1a_2a_3$. This is exactly 
the result for $SU(3)$ with one hypermultiplet in the
fundamental representation obtained for example in \cite{mas1,
mas2}, also with a change in the quantum scale $\La^4\rightarrow
\frac{\La^4}{4}$.

\section{Conclusions}

In the present paper we have studied the form of the prepotential 
of ${\cal N}=2$ supersymmetric gauge theories with gauge group $SU(N_1)\times
SU(N_2)$ with a hypermultiplet in the bifundamental representation and
matter in the fundamental representation of both gauge groups. 
The Seiberg--Witten curves for those theories are
non--hyperelliptic curves derived from M--theory considerations. 
In the first place we calculate the
logarithmic derivatives of the prepotential with respect to the
quantum scales of both gauge groups. With the help of the Riemann 
bilinear relations we express this logarithmic derivative in terms of 
the moduli of the Seiberg--Witten curve. In fact we find that  
this logarithmic derivative of the prepotential with respect to the quantum 
scales $\La_1$ and $\La_2$ shows a non-trivial mixing between both groups 
order parameters, as is expected in the presence of a hypermultiplet in the 
bifundamental representation. As an application we develop a 
method to compute recursively the instanton corrections to the prepotential 
with the help of the previously calculated equation. 
Using this, we 
find that we can compute the instanton corrections recursively in a
straightforward way. This improves the existing method developed in
\cite{sch1}--\cite{sch4}. In that references they compute the first
instanton correction calculating explicitly the dual order parameters
and integrating them. That method gets very complicated for higher
corrections, so does not allow one to go further. The method
described here is much simpler as we avoid the calculation of the dual 
periods. 
We also extend the method to compute the instanton corrections to 
the non-hyperelliptic curves obtained for $SU(N)$ theories with 
matter in the symmetric and antisymmetric representation, finding also
that it allows us to compute recursively the instanton corrections to the
prepotential.

It is important to note that for the case of pure $SU(N)$, or with
matter in the fundamental representation, the Seiberg--Witten curves
were originally calculated using just field theory
considerations. Therefore the comparison of the results obtained using
the Seiberg--Witten curves with the ones obtained from a
microscopic calculation (or from other 
alternative methods, like the one presented in \cite{nekrasov}) are 
seen as tests of the Seiberg--Witten
approach. Nevertheless, the Seiberg--Witten curves for supersymmetric 
gauge theories with product gauge groups, or matter in the symmetric or 
antisymmetric representation of the gauge group, are non--hyperelliptic curves 
 derived just from M--theory considerations. Therefore, when 
microscopic calculations will become available for these theories, 
the comparison of results is not just a test of the Seiberg--Witten approach 
but also from the M--theory
considerations used to derive the Seiberg--Witten curves.

Also we must point out that the recent papers of Dijkgraaf and 
Vafa \cite{vafa1} have
attracted a new attention to the subject of Seiberg--Witten
theory. Their work indicates that several non--perturbative results
in supersymmetric gauge theories can be obtained by means of
perturbative calculations using auxiliary matrix models. It would be
interesting to see if the results obtained here can be reproduced 
from the matrix model approach.

\section*{Acknowledgments}

The author would like to thank J. Mas for many
useful discussions and for the careful reading of the manuscript. This
work has been supported by Fundaci\'on Ram\'on Areces.


\end{document}